\documentclass[onecolumn,showpacs,preprintnumbers,amsmath,amssymb]{revtex4}
\usepackage[english]{babel}
\usepackage{graphics}
\usepackage{epsf}
\usepackage{psfig}
\usepackage{dcolumn}
\usepackage{bm}
\begin{document}

\title{Quantum Energy in a Vibrating Cavity}

\author{Pawe\l{} W\c{e}grzyn}
\affiliation{ Marian Smoluchowski Institute of Physics,
Jagellonian University, Reymonta 4, 30-059 Cracow, Poland}
\email{wegrzyn@th.if.uj.edu.pl}

\begin{abstract}
We discuss the quantized  field inside a general one-dimensional
cavity system. We recognize the $SL(2,R)$ symmetry being the
remainder of the conformal group. The explanation of the lack of
the resonance production for the fundamental frequency is given
and the asymptotic behavior of the cavity system is properly
described.
\end{abstract}


\maketitle

The quantized massless scalar field in one spatial dimension
restricted to live in a finite but periodically oscillating cavity
was studied in many papers. It results in the generation of
particles due to the instability of the vacuum state in the
presence of time-dependent boundaries. This subject has been
reviewed recently in \cite{review}. From experimental point of
view,  the most interesting aspect is the instability of the
system triggered by the parametric opto-mechanical resonance. In
these cases  the total energy of the field is exponentially
growing with time.

Investigations concerning time-dependent cavity systems, referred
usually as dynamical Casimir effect or motion-induced radiation,
have been performed mainly in one spatial dimension. Our paper is
a supplement to the subject and we reveal the role of the remnant
$SL(2,R)$ symmetry in such investigations. Relying on the
symmetry, we derive general  formulas for quantum energies and
numbers of produced particles and we consider basic examples of
cavity systems in one spatial dimension.

The standard model consists in the massless scalar field $A(x,t)$
which satisfies the classical wave equation with Dirichlet
boundary conditions. The left cavity wall is  fixed at position
$x=0$, while the right wall is oscillating with some prescribed
trajectory $L(t)$. It is reasonable to assume that the cavity
oscillates during some finite time period $0<t<T$. For $t<0$ and
$t>T$, the cavity is static and its length is $L$. The complete
set of basic classical solutions for this problem was found in
\cite{moore}:
\begin{equation}
A_k(x,t)=N_k \left[ \exp{(-i \omega_k R(t+x) )} - \exp{(-i
\omega_k R(t-x) )} \right] \ ,
\end{equation}
where the resonance frequencies are $\omega_k=k \omega$, with
$\omega = \omega_1= \pi/L$, the normalization factor is
$N_k=i/(4\pi k)^{-1/2}$ and the phase function is to be derived
from the Moore equation:
\begin{equation}\label{me}
R(t+L(t))-R(t-L(t))=2L \ .
\end{equation}
It will be more useful in later applications to express the above
equation in the following way:
\begin{equation}\label{me2}
\frac{1}{2L} \int_{t-L(t)}^{t+L(t)} \, d\tau \, \dot{R}(\tau)=1
 \ .
\end{equation}
The quantization of the field follows the development given in
\cite{moore, fulling}, and we recall the results obtained there
(see also \cite{wegrzyn2}). The new aspect of our treatment is the
detailed discussion of the quantum energy. Therefore, we skip the
quantization procedure and carry over main  results from previous
papers, and we focus on the output resulting in the formulas for
the total energy and the number of produced particles.

The vacuum expectation value for the energy density inside the
cavity can be presented in the following form:
\begin{equation}
\langle    T_{00}(x,t) \rangle = \langle 1/2(\partial_t
A)^2+1/2(\partial_x A)^2 \rangle =\varrho(t+x)+\varrho(t-x) \ ,
\end{equation}
where the profile function for left- and right-movers is defined
by:
\begin{equation}\label{profile}
\varrho(\tau)=-\frac{\pi}{48 L^2}
\dot{R}^2(\tau)-\frac{1}{24\pi}S[R](\tau) \ .
\end{equation}
The second term is responsible for particle production and it is
defined by the Schwartz derivative:
\begin{equation}
S[R]=\frac{\stackrel{...}{R}}{\dot{R}}-\frac{3}{2}
\left(\frac{\stackrel{..}{R}}{\dot{R}}\right)^2 \ .
\end{equation}
The formula for the total energy can be presented in the following
form:
\begin{equation}\label{te}
E(t)=\int_0^{L(t)} \ dx \ \langle T_{00}(x,t) \rangle
=\int_{t-L(t)}^{t+L(t)} \ d\tau \ \varrho(\tau) \ .
\end{equation}
Obviously, if the cavity ends the oscillations and returns to its
static state, then the total energy reaches its final value.
Therefore, for all $t>T$ we have $E(t)=E(T)$. However, the energy
density is still changing with time. We can observe travelling
wave packets which bounce between cavity walls. For $t>T$, it is
meaningful to define the Bogolubov coefficients which are
eventually time-independent:
\begin{equation}
\beta_{kl}=-\frac{1}{2L}\sqrt{\frac{l}{k}}\int^{t+L}_{t-L} \ d\tau
\ e^{-i\omega_k R(\tau)-i\omega_l \tau} \ .
\end{equation}
The number of photons in $\it k$-th mode and the total number of
photons produced in the oscillating cavity are given by:
\begin{equation}
n_k=\sum^{\infty}_{l=1} |\beta_{lk}|^2 \ \ , \ \
N=\sum^{\infty}_{k=1} n_k
\end{equation}

Let us now begin with the analysis of the above formulas. First,
we consider the trivial static case $L(t)=L$. The Moore equation
Eq.(\ref{me}) possesses many solutions, which correspond to
different choices of the basis in the quantum space of field
states. A particular solution is the  identity function $R(z)=z$
which provides us with the standard basis composed of standing
waves. In the static case, the composition of any solutions of
Eq.(\ref{me}) gives also a solution. If we add that the phase
function must be invertible, we see that the set of solutions of
Eq.(\ref{me}) forms a group. Finally, in any basis the
calculations in the static case lead to the same conclusion that
the number of produced particles is zero: $N=0$, and the total
energy equals the Casimir zero energy: $E=-\pi/24L=-\omega/24$.

Next, we will examine the dynamical case.  Looking at
Eq.(\ref{profile}), we cannot guess whether the total energy is
bounded from below. We should prove that to provide the ultimate
justification of the quantization of the model. The contribution
from the first term of Eq.(\ref{profile}) is always negative. It
is easy to derive from Eq.(\ref{me2}) that this contribution is
even smaller than the static Casimir energy for the instant value
of the distance between mirrors:
\begin{equation}
-\frac{\pi}{48 L^2} \int_{t-L(t)}^{t+L(t)} \ d\tau \
\dot{R}^2(\tau) \leq - \frac{\pi}{24} \frac{1}{L(t)} \ .
\end{equation}
If we refer to the terminology used  in literature, this negative
contribution may be called {\it sub-Casimir energy}. Depending on
the specific cavity motion, this negative energy part can be
arbitrarily large. It may be freely large even if the distance
between cavity walls is kept big enough, so that the cavity length
is never smaller than some fixed value.

In order to simplify the analysis of the total energy in its full
extent, we introduce the following function:
\begin{equation}\label{sub}
T(\tau) = \left( \frac{d}{d\tau} R^{-1}(\tau) \right)^{-1/2} \ .
\end{equation}
A nice feature of the above substitution is that it leads to the
quadratic form in the integrand of Eq.(\ref{te}):
\begin{equation}\label{te22}
E= \frac{1}{12\pi} \int_{t'-L}^{t'+L} \ d\tau \ \left(
\dot{T}^2-\frac{\omega^2}{2}T^2\right) \ ,
\end{equation}
where $t'=R(t-L)+L=R(t+L)-L$ and the Moore equation Eq.(\ref{me2})
is now equivalent to:
\begin{equation}\label{me3}
\frac{1}{2L} \int_{t'-L}^{t'+L} \, d\tau \, \frac{1}{T^2(\tau)} =
1 \ .
\end{equation}
We examine the cavity in its final stage for $t>T$. It is
straightforward to find the stationary points of the action
Eq.(\ref{te22}) with the constraint Eq.(\ref{me3}). They
correspond to the following solutions:
\begin{equation}
T_{min}^2(\tau)=
\frac{1}{2}(A^2+B^2+C^2+D^2)+\frac{1}{2}(A^2+B^2-C^2-D^2)
\cos{(\omega\tau)}-(AC+BD)\sin{(\omega\tau)} \ ,
\end{equation}
where the integration constants are real numbers that satisfy the
condition $AD-BC=1$. Using Eq.(\ref{sub}), we can evaluate the
phase functions for cavity motions that minimizes the total
energy:
\begin{equation}\label{min}
R_{min}(\tau)=\frac{2}{\omega}  \arctan{\left( \sigma\left(
\tan{\frac{\omega \tau}{2}} \right) \right)} \ ,
\end{equation}
where the consecutive branches  of the multivalued function arctan
should be properly chosen to have the phase function increasing.
We have introduced the inner function:
\begin{equation}
\sigma(\tau)=\frac{A\tau+B}{C\tau+D}  \ .
\end{equation}
It is easy to verify that for all solutions Eq.(\ref{min}) the
number of produced photons is just zero $N=0$ and the total energy
matches the static Casimir value $E=-\omega/24$. The set of
minimal solutions forms a representation of $SL(2,R)$ group. The
inverse of the phase function is given by:
\begin{equation}
R_{min}^{-1}(\tau)=\frac{2}{\omega}  \arctan{\left(
\sigma^{-1}\left( \tan{\frac{\omega \tau}{2}} \right) \right)} \ ,
\end{equation}
where $\sigma^{-1}(\tau)=(D\tau-B)/(-C\tau+A)$. The existence of
$SL(2,R)$ symmetry is related to conformal transformations that
leave space boundaries unchanged. These transformations are just:
\begin{equation}\label{sym}
t\pm x\longrightarrow R_{min}(t\pm x) \ .
\end{equation}
To consider infinitesimal transformations, we put $A=1+a$, $B=b$,
$C=c$ and $D=1-a$,  where parameters $a,b,c$ are assumed to be
small. We obtain:
\begin{eqnarray}
  \omega t &\longrightarrow& \omega t+(b-c)+
  (b+c)\cos{(\omega t)} \cos{(\omega x)}+ 2a\sin{(\omega t)}
  \cos{(\omega x)} \nonumber \\
  \omega x &\longrightarrow& \omega x-
  (b+c)\sin{(\omega t)} \sin{(\omega x)}+ 2a\cos{(\omega t)}
  \sin{(\omega x)}
\end{eqnarray}
A group of $SL(2,R)$ transformations is composed of time
translations and periodic deformations of coordinates with the
fundamental frequency $\omega$. For any cavity motion, it is easy
to check that for final times $t>T$ the transformed phase function
$R\circ R_{min}$ corresponds to the same energy density
Eq.(\ref{profile}) and the same number of particles $N$ as those
obtained using the function $R$.

The existence of $SL(2,R)$ symmetry suggests us that it is
convenient to represent the resonant solutions of the Moore
equation in the following form:
\begin{equation}\label{subst}
R(\tau)=\frac{2}{\omega_k}  \arctan{\left( \sigma_j\left(
\tan{\frac{\omega_k \tau}{2}} \right) \right)} \ , \ \ \ \ \ \
j=1,2,...,k \ ,
\end{equation}
where $\sigma_j(\tau)$ are now  arbitrary increasing functions.
The period interval $2L$ that appears in integral formulas is now
divided into $k$ equal pieces.  In general, the inner function
$\sigma$ may differ in the subintervals and it justifies the
subscript $j$. The profile function of the energy density
Eq.(\ref{profile}) is given by:
\begin{equation}\label{prof2}
\varrho(\tau)=-\frac{k^2\omega^2}{48\pi} \left( 1+(1+\upsilon^2)^2
S[\sigma_j](\upsilon) \right) +\frac{(k^2-1)\omega^2}{48\pi}
\frac{(1+\upsilon^2)^2
\dot{\sigma_j}^2(\upsilon)}{(1+\sigma_j^2(\upsilon))^2} \ ,
\end{equation}
where $\upsilon=\tan{(\omega_k \tau/2)}$. The total energy is
equal to:
\begin{equation}\label{te2}
E=-k^2\frac{\omega}{24} \left( 1 + \frac{1}{k\pi} \sum_j
\int_{-\infty}^{+\infty} d\upsilon
(1+\upsilon^2)S[\sigma_j](\upsilon)
\right)+\frac{(k^2-1)\omega}{24k\pi} \sum_j
\int_{-\infty}^{+\infty} d\upsilon \frac{(1+\upsilon^2)
\dot{\sigma_j}^2(\upsilon)}{(1+\sigma_j^2(\upsilon))^2} \ ,
\end{equation}
and the Bogolubov coefficients can be calculated from:
\begin{equation}\label{bc2}
\beta_{kl}=\frac{(-1)^{k+l+1}}{\pi} \sqrt{\frac{l}{k}} \sum_j
\int_{-\infty}^{+\infty} \frac{d\upsilon}{1+\upsilon^2} \left(
\frac{\upsilon+i}{\upsilon-i} \right)^{l}  \left(
\frac{\sigma_j(\upsilon)+i}{\sigma_j(\upsilon)-i} \right)^{k} \ .
\end{equation}
We apply the above description and formulas to consider several
examples of models considered in literature. Let us begin with the
model of sinusoidal cavity oscillations:
\begin{equation}\label{sin}
    L(t)=L+\Delta L \sin{(\omega_k t)} \ .
\end{equation}
Obviously, we are to assume that $\Delta L<L$ and $\omega_k \Delta
L<1$. The asymptotic solution for phase function was found in
\cite{dodklim, Dodonov2} in the following form:
\begin{equation}\label{po}
R(\tau)=\tau-\frac{2}{\omega_k} {\rm Im} \left( \ln{\left[
1+\zeta+e^{i\omega_k\tau}(1-\zeta) \right]} \right) \ ,
\end{equation}
where $\zeta=\exp{[(-1)^{k+1}\omega_k \tau \Delta L /L]}$. It is
straightforward to rewrite this solution as:
\begin{equation}
R(\tau)=\frac{2}{\omega_k}  \arctan{\left( \zeta
\tan{\frac{\omega_k \tau}{2}}  \right)}   \ ,
\end{equation}
or we can simply state that $\sigma_j(\upsilon)=\zeta \upsilon$.
For large times $T$, we can consider $\zeta$ to be constant and
calculate easily the total energy from Eq.(\ref{te2}):
\begin{equation}
E= -k^2 \frac{\omega}{24} +\frac{(k^2-1)\omega}{24}
\cosh{(\omega_k T \Delta L/L)} \ .
\end{equation}
This result agrees with \cite{dodklim, Dodonov2,bordag}. The lack
of the resonance production of particles for the lowest frequency
$\omega_1=\omega$ is due to the $SL(2,R)$ symmetry.

Some exact analytical solution corresponding to a vibrating cavity
system was presented in \cite{law} for the resonance channel $k=2$
and generalized in \cite{wu} for higher resonances. The solutions
correspond to a family of wall motions:
\begin{equation}\label{lawwu}
    L(t)=L+\frac{1}{\omega_k} \left\{ \arcsin{\left[
    \sin{\frac{\omega_k \Delta L}{2}} \cos{(\omega_k t)} \right] -
\frac{\omega_k \Delta L}{2} } \right\}
 \ . \end{equation}
The solutions for corresponding phase functions can be presented
in the form Eq.(\ref{subst}) with
\begin{equation}
\sigma_j(\upsilon)=\upsilon/[1-(T/L-1)\upsilon\tan{(\omega_k\Delta
L/2)}] \ , \end{equation} for $k $ even and
\begin{equation}
\sigma_j(\upsilon)=\upsilon+(T/L-1)\tan{(\omega_k\Delta L/2)} \ ,
\end{equation}
for $k $ odd. Again, the Schwartz derivatives $S[\sigma_j]$
vanish. It is straightforward to obtain the energy density and the
Bogolubov coefficients. In this case, the energy happens to grow
quadratically with time.

Using Eq.(\ref{te2}) and Eq.(\ref{bc2}), we can prove for our
examples of cavity systems that the standard relation between the
total energy and the number of particles in particular modes
holds:
\begin{equation}
E=-\frac{\omega}{24} + \sum_k n_k \omega_ k \ .
\end{equation}

In conclusion, we have demonstrated that the quantum energy in a
time dependent cavity system is well defined. The total energy is
always greater than the Casimir energy of the corresponding static
cavity system. However, the local expectation values of the energy
density may take  arbitrary large negative values. Unlike other
previous studies on the same subject, we show the important role
of $SL(2,R)$ symmetry in the analysis of cavity systems. This
symmetry is a remainder of the conformal symmetry. It helps to
explain the lack of the mechanism of particle production for the
lowest, fundamental resonance frequency. We also described the
asymptotic behavior of the resonant cavity system.

%\newpage
%\psfig{file=r1.eps, width=16cm,height=22cm}


\begin{thebibliography}{99}

\bibitem{review} V.V.Dodonov, Modern Nonlinear Optics, Part 1,
p.309-394
 (Advances in Chemical Physics, Volume 119, ed. by M.W.Evans, Wiley, N.Y.,
 2001).
\bibitem{moore}{G.T. Moore. J. Math. Phys. {\bf 11}, 2679
(1970).}
\bibitem{fulling} S.A. Fulling, C.W. Davies, Proc. R. Soc. Lond. A {\bf
348}, 393 (1976).
\bibitem{wegrzyn2} P. W\c{e}grzyn, T. R\'og, Acta Phys. Polon.
{\bf B 34}, 3887 (2003).
\bibitem{dodklim} V.V.Dodonov, A.B.Klimov, Phys. Lett. {\bf A
167}, 309 (1992).
\bibitem{Dodonov2} V.V. Dodonov, A.B. Klimov, D.E. Nikonov, J.Math.Phys.
{\bf34}, 2742 (1993).
\bibitem{bordag} M.Bordag, U.Mohideen, V.M.Mostepanenko, Physics Reports {\bf
353}, 1-205 (2001).
\bibitem{law} C.K.Law, Phys. Rev. Lett. {\bf 73}, 1931 (1994).
\bibitem{wu} Y.Wu, K.W.Chan, M.C.Chu, P.T.Leung, Phys.Rev. {\bf A 59},
1662 (1999).

%\bibitem{berges} J. Berges, J. Serreau, Phys. Rev. Lett. {\bf 91},
%111601 (2003).
%\bibitem{lambrecht} A.Lambrecht, M.T.Jaeckel, S.Reynaud,
%Phys. Rev. Lett. {\bf 77}, 615 (1996).
%\bibitem{bordag} M.Bordag, U.Mohideen, V.M.Mostepanenko, Physics Reports {\bf
%353}, 1-205 (2001).
%\bibitem{meplan} O. M\'{e}plan, C. Gignoux, Phys. Rev. Lett. {\bf 76}, 408 (1996).
%\bibitem{nagatani} Y.Nagatani, K.Shigetomi, Phys. Rev. {\bf A 62},
%022117 (2000).

%bibitem{dittrich} J.Dittrich, P.Duclos, P.\u{S}eba, Phys. Rev.
%{\bf E 49}, 3535 (1994).
%\bibitem{wegrzyn} P. W\c{e}grzyn, T. R\'og, Acta Phys. Polon. {\bf
%32}, 129 (2001).
%\bibitem{cole} C.K. Cole, W.C. Schieve, Phys.Rev. {\bf A 52}, 4405
%(1995).
%\bibitem{arnold} V. I. Arnold, {\it Mathematical Methods Of
%Classical Mechanics} (Springer, New York, 1978), Sec. 25 and Sec.
%42.
%\bibitem{lambrecht2} A.Lambrecht, M.T.Jaeckel, S.Reynaud,
%Europhys. Lett. {\bf 43}, 147 (1998).
%\bibitem{Dodonov1} V.V. Dodonov, A.B. Klimov, D.E. Nikonov, Phys.
%Lett A {\bf149}, 225 (1990).
%\bibitem{Jaeckel} M.T.Jaeckel, S.Reynaud, J. Phys. I (France) {\bf
%2}, 149 (1992).

%bibitem{Law2} C.K.Law, Phys. Rev. A {\bf 49}, 433 (1994).
%\bibitem{Dodonov3} V.V. Dodonov, A.B. Klimov, Phys. Rev. A {\bf 53},
%2664 (1996).

%\bibitem{Dodonov4} V.V.Dodonov, J.Phys. A: Math. Gen. {\bf31},9835
%(1998).
%\bibitem{Dalvit1} D.A. Dalvit, F.D. Mazzitelli, Phys. Rev. A {\bf57}, 2113
%(1998).
%\bibitem{Dalvit2} D. A. Dalvit, F.D. Mazitelli, Phys. Rev. A {\bf59}, 3049
%(1999) .
%\bibitem{Walker} W.R.Walker, Phys. Rev. D {\bf31}, 767 (1985).

\end{thebibliography}
\end{document}